\documentclass[10pt,conference]{IEEEtran}
\usepackage{amsmath, amssymb, amsfonts}
\usepackage{mathtools}
\usepackage{bm}
\usepackage{graphicx}
\usepackage{cite}
\usepackage{color}
\usepackage{booktabs}
\usepackage{url}
\usepackage{hyperref}
\usepackage{stfloats}

\usepackage{tabularx}
 \newcolumntype{L}{>{\raggedright\arraybackslash}X}
\usepackage{microtype}
\usepackage{enumitem}
\usepackage{float}
\usepackage{textcomp}
\usepackage{makecell}
\usepackage{svg}
\usepackage{multirow}
\usepackage{threeparttable}  

\IEEEoverridecommandlockouts

\def\BibTeX{{\rm B\kern-.05em{\sc i\kern-.025em b}\kern-.08em
    T\kern-.1667em\lower.7ex\hbox{E}\kern-.125emX}}
\usepackage{balance}

\begin{document}
	
\title{Online Learning Extreme Learning Machine with Low-Complexity Predictive Plasticity Rule and FPGA Implementation}
	
\author{Zhenya Zang, Xingda Li, and David Day Uei Li%
\thanks{This work is supported by the EPSRC (EP/T00097X/1), the Quantum Technology Hub in Quantum Imaging (QuantiC), and the University of Strathclyde. Xingda Li also acknowledges support from the China Scholarship Council. The authors would also like to acknowledge the
support from Xilinx for donating the FPGA. }%
\thanks{Zhenya Zang, Xingda Li, and David Day Uei Li are with the Department of Biomedical Engineering, University of Strathclyde, Glasgow, UK. (e-mail: zhenya.zang@strath.ac.uk, xingda.li@strath.ac.uk, david.li@strath.ac.uk).}
}

\maketitle
	
\begin{abstract}
    We propose a simplified, biologically inspired predictive local learning rule that eliminates the need for global backpropagation in conventional neural networks and membrane integration in event-based training. Weight updates are triggered only on prediction errors and are performed using sparse, binary-driven vector additions. We integrate this rule into an extreme learning machine (ELM), replacing the conventional computationally intensive matrix inversion. Compared to standard ELM, our approach reduces the complexity of the training from $\mathcal{O}(M^3)$ to $\mathcal{O}(M)$, in terms of $M$ nodes in the hidden layer, while maintaining comparable accuracy (within $3.6\%$ and $2.0\%$ degradation on training and test datasets, respectively). We demonstrate an FPGA implementation and compare it with existing studies, showing significant reductions in computational and memory requirements. This design demonstrates strong potential for energy-efficient online learning on low-cost edge devices.
\end{abstract}

\begin{IEEEkeywords}
    Bio-inspired learning, Online learning, Reconfigurable hardware
\end{IEEEkeywords}
	
\section{Introduction}
\IEEEPARstart{O}{nline} learning is essential in modern machine learning (ML) systems that operate under nonstationary data distributions, dynamic environments, or real-time constraints. Although modern deep learning methods have achieved state-of-the-art performance in various domains, they are primarily based on gradient-based optimisation and global error backpropagation \cite{lecun2015deep, rumelhart1986learning}. Although effective in batch learning, these approaches are inherently inefficient and ill-suited for real-time online adaptation. Alternative paradigms have emerged to address these challenges. Neuromorphic computing \cite{roy2019towards}, for example, emulates the neural activities of the brain using event-driven parallel hardware to achieve highly efficient computation, making it ideal for edge AI and real-time processing. Representative digital neuromorphic implementations have demonstrated high scalability \cite{Scalable2020}, enabling biologically inspired cognitive architectures \cite{BiCoSS2022}, cerebellar network models \cite{CerebelluMorphic2020}, and fault-tolerant context-dependent learning frameworks \cite{Fault_Tolerant2022}. Representative training rules, such as spike-timing-dependent plasticity (STDP) \cite{STDP2008} and predictive learning rules (PLR)~\cite{PLR_NC_2023} enable local learning at the synapse level. These rules assign credit based on temporal correlations or predictive errors, aligning well with streaming data and hardware-locality constraints. Another promising direction involves backpropagation-free methods that eliminate computationally intensive training, such as Extreme Learning Machine (ELM)~\cite{ELM, OS_ELM}, and Random Vector Functional Link~\cite{RVFL2023}. ELM, featuring fast training and a simple architecture, provides an efficient alternative to conventional neural networks by projecting input data through a fixed random layer and solving output weights via a closed-form least-squares solution, thereby avoiding iterative gradient updates. However, standard ELMs assume full access to training data and require costly matrix inversion, limiting their suitability for real-time and low-power inference. To overcome these issues, several online ELM variants~\cite{OS_ELM, OS_ELM_FORGOTTEN, ELM_ONLINE_HLS} have been proposed, incorporating recursive least-squares updates and forgetting mechanisms for incremental learning.

In this work, we propose a compact learning architecture, simplified PLR-ELM (SPLR-ELM), which combines the fast inference capability of ELMs with the low complexity of PLR. The proposed method replaces the global least-squares solution with a local winner-take-all (WTA) predictive update based on an error-driven learning rule, eliminating the need for eligibility traces, membrane potential integration, and global backpropagation. This design enables an efficient online update path that relies only on binary vector additions. We demonstrate a fixed-point (FXP) FPGA implementation using binary hidden activations and multiplier-free update logic. Unlike conventional ELMs or biologically inspired models, SPLR-ELM provides a scalable on-device learning solution for edge scenarios constrained by energy, memory, and latency. This is the first work to unify ELM and simplified PLR within an efficient online learning framework and validate its practicality through hardware realisation, integrating biologically inspired plasticity with efficient hardware design for real-time ML systems.

\section{Theories}

\begin{table*}[h!]
  \centering
  \caption{Conceptual and Methodological Comparison of Training from ELM, PLR, and the Proposed SPLR}
  \begingroup
    \footnotesize
    \scriptsize  
    \renewcommand{\arraystretch}{0.55}
  \begin{tabularx}{\linewidth}{@{} L L L L @{}}
    \toprule
    \textbf{Feature} & \textbf{Standard ELM}~\cite{ELM} & \textbf{Original PLR}~\cite{PLR_NC_2023} & \textbf{Proposed SPLR} \\
    \midrule
    Loss Function & Minimises a global least-squares & Minimises a predictive loss & Minimises a WTA loss \\
    \midrule
    Learning Rule & $\mathbf{W}^{*} = \mathbf{H}^\dagger \mathbf{T} = (\mathbf{H}^\top \mathbf{H})^{-1}\mathbf{H}^\top\mathbf{T}$ & 
    $w_t = w_{t-1} + \eta(\epsilon_t v_{t-1} + \varepsilon_t p_{t-1})$ &
    $w = w + \eta h;\; w = w - \eta h$ \\
    \midrule
    Learning Scale & Global & Neuron-wise, temporal \& error driven & Neuron-wise \& error driven \\
    \midrule
    Trace & N.A. & Eligibility traces & Removed for simplicity \\
    \midrule
    Computational Complexity & $\mathcal{O}(M^3)$ (matrix inversion) & $\mathcal{O}(M)$ per time step & $\mathcal{O}(M)$ when misclassified \\
    \midrule
    Hardware Suitability & Low, high complexity & Low, primarily for spiking systems & High, only additions in weight update \\
    \bottomrule
  \end{tabularx}
  \endgroup
  \label{tab:learning_compare}
\end{table*}

\subsection{Closed-Form Solution of ELM}
In ELM \cite{ELM}, input-to-hidden weights are randomly initialised and fixed, while only hidden-to-output weights are learned. For an input $\mathbf{x} \in \mathbb{R}^D$, the hidden activation is $\mathbf{h} = \sigma(\mathbf{W}_{\text{in}} \mathbf{x} + \mathbf{b})$, where $\sigma(\cdot)$ is a nonlinear function, and $\mathbf{W}_{\text{in}} \in \mathbb{R}^{M \times D}$ and $\mathbf{b} \in \mathbb{R}^M$ are random and fixed. Given $N$ samples, hidden responses form $\mathbf{H} \in \mathbb{R}^{N \times M}$ and targets $\mathbf{T} \in \mathbb{R}^{N \times C}$; trainable weights $\mathbf{W} \in \mathbb{R}^{M \times C}$ are then estimated by minimizing the least-squares error.
\begin{equation}
    \min_{\mathbf{W}} \left\| \mathbf{H}\mathbf{W} - \mathbf{T} \right\|_2^2.
\end{equation}
The learning rule of ELM is summarised in Table~\ref{tab:learning_compare}. Several online ELM variants~\cite{ELM_ONLINE_HLS, OS_ELM, OS_ELM_FORGOTTEN} adopt incremental updates to reduce computational cost and improve on-chip efficiency. However, they still require an initial batch training phase involving the matrix inversion $(\mathbf{H}_0^\top \mathbf{H}_0)^{-1}$, where $\mathbf{H}_0 \in \mathbb{R}^{N_0 \times M}$ and $N_0$ denotes the number of samples in the initial batch. Neuromorphic methods, in contrast, avoid large-scale matrix operations during training and inference with minimal accuracy loss. Following this principle, the proposed SPLR-ELM replaces global learning with local error-driven predictive updates without spike-train encoding, effectively balancing accuracy and computational cost for real-time, low-power deployment.

\subsection{Simplified Predictive Learning Rule}
PLR \cite{PLR_NC_2023} is a biologically inspired local learning rule that enables single neurons to predict future events by capturing the temporal structure of synaptic inputs. It strengthens synapses that best anticipate others, giving rise to sequence learning, anticipatory spiking, and STDP-like behavior. PLR has been effectively used to train spiking neural networks (SNNs) and has been implemented on FPGA \cite{FPGA_PLR_SNN_FPGA}, demonstrating both the learning efficiency and hardware suitability. The original weight updated role is presented in Table \ref{tab:learning_compare},
where $\eta$ is the learning rate (LR), $\epsilon_t = x_t - v_{t-1} w_{t-1}$ denotes the prediction error, $\varepsilon_t = \epsilon_t w_{t-1}$ represents the global modulatory signal, and $p_t = p_{t-1} e^{-\frac{t}{\tau_m}} + x_t$ is the eligibility trace. This formulation allows individual neurons in an SNN to learn long-timescale temporal dependencies and shift their responses toward predictive inputs, aligning with experimental observations of STDP. The weight update rule requires continuous tracking of synaptic state variables across time, such as eligibility traces $p_t$ and membrane potentials $v_t$, involving floating-point (FLP) multiplications. While biologically plausible, such complexity introduces substantial computational overhead, particularly on hardware-constrained platforms. 
    
To address this challenge, our proposed SPLR replaces continuous dynamics with a discrete, error-driven mechanism: HL outputs are one-timestamp, binary spikes $\mathbf{h} \in \{0,1\}^M$, and output predictions are made through a linear readout $\mathbf{o} = \mathbf{h}^\top \mathbf{W}$, where
\begin{equation}
\mathbf{h} = \Theta\left(\mathbf{W}_{\text{in}} \mathbf{x} + \mathbf{b}\right) \in \{0,1\}^M.
\label{eq:hardset}
\end{equation}
Unlike prior work~\cite{FPGA_PLR_SNN_FPGA}, which uses leaky integrate-and-fire (LIF) models in the HL, our SPLR-ELM receives normalized real-valued input $\mathbf{x} \in \mathbb{R}^D$ and projects it to binary hidden states using a Heaviside step function activation $\Theta(\cdot)$ with a fixed threshold. This threshold can be replaced by zero-centering the projection $\mathbf{W}_{\text{in}} \mathbf{x} + \mathbf{b}$, but with a $\sim$3\% drop in classification accuracy. The update rule shown in Table  \ref{tab:existing_fpga_compare} is applied only when the prediction is incorrect. $\mathbf{W}$ is adjusted by reinforcing the correct output and suppressing the incorrect one. This rule eliminates the need for eligibility traces, membrane potential integration, and FLP multiplications by integer-based vector additions or subtractions. Weight values are clipped to a bounded range to avoid divergence:
\begin{equation}
    w \leftarrow \min\left(\max(w, -w_{\max}), w_{\max}\right).
\end{equation} 
To prove the feasibility of the conventional gradient-based training perspective, SPLR-ELM’s weight update can be interpreted as the gradient of a loss function based on WTA behavior
\begin{equation}
    \mathcal{L} = \frac{1}{2} (o_{\hat{y}} - o_y)^2.
\end{equation}
The gradients of the loss function with respect to the output weights are
\begin{align}
    \frac{\partial \mathcal{L}}{\partial W_{i\hat{y}}}
    &= \frac{\partial \mathcal{L}}{\partial o_{\hat{y}}} \cdot \frac{\partial o_{\hat{y}}}{\partial W_{i\hat{y}}}
    = (o_{\hat{y}} - o_y) \cdot h_i, \\
    \frac{\partial \mathcal{L}}{\partial W_{iy}}
    &= \frac{\partial \mathcal{L}}{\partial o_y} \cdot \frac{\partial o_y}{\partial W_{iy}}
    = - (o_{\hat{y}} - o_y) \cdot h_i,
\end{align} 
which are hardware-friendly, since $h_i$ is binary and the updates involve only addition or subtraction operations.
\begin{figure}[!t]
    \centering
    \includegraphics[width=0.9\linewidth]{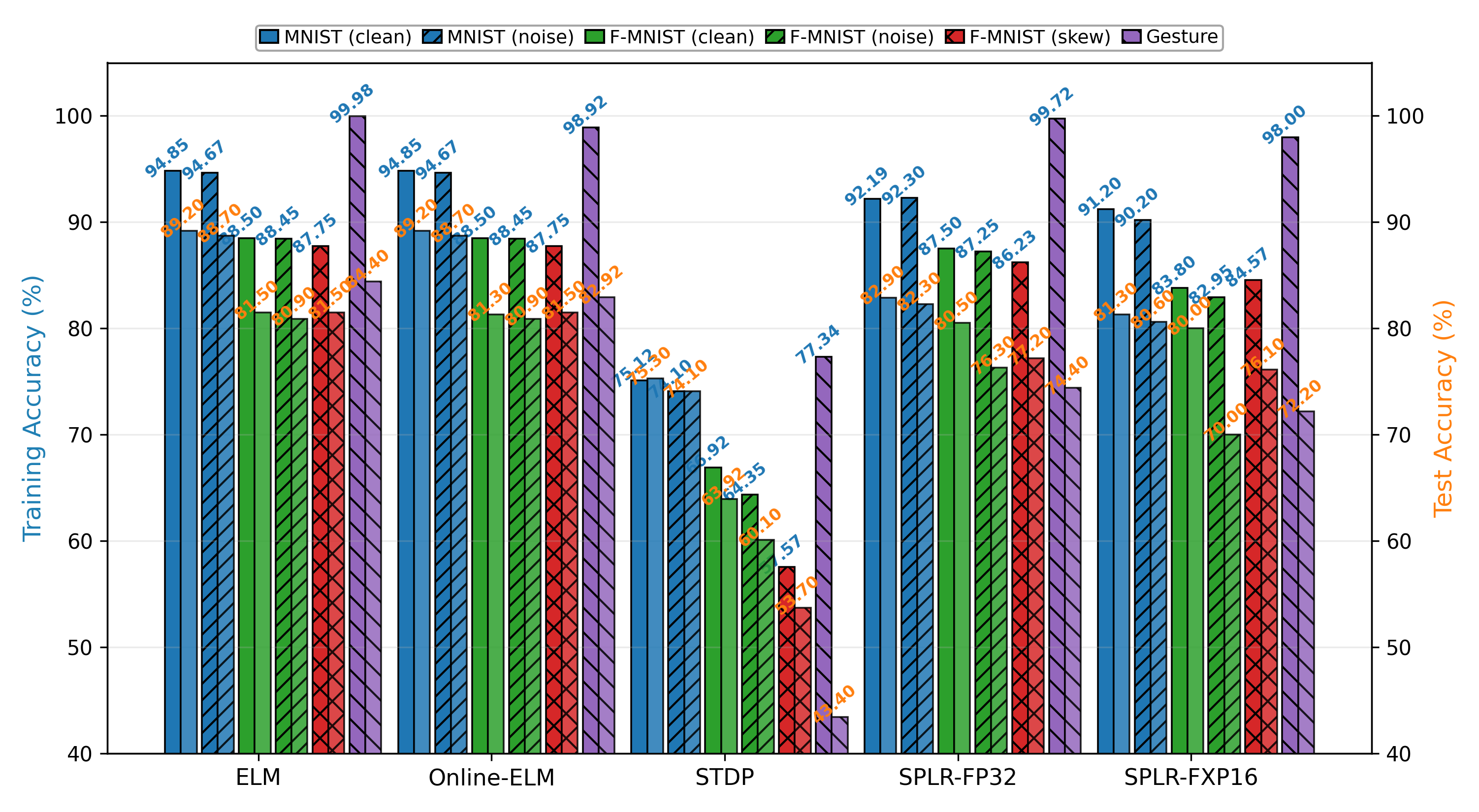}
    \caption{Comparison of training and test accuracy with different learning paradigms on the MNIST and Fashion-MNIST datasets.}
    \label{fig:Bar_chart}
\end{figure}
Using a fixed LR $\eta$ and assuming hard WTA behaviour, the update rule can be discretised to
\begin{equation}
    \Delta W_{iy} = +\eta h_i, \quad
    \Delta W_{i\hat{y}} = -\eta h_i \quad \text{if } \hat{y} \ne y,
        \label{eq:WU}
\end{equation}
which is equivalent to the rule shown in Table \ref{tab:learning_compare}. Compared with the intensive one-time matrix inversion in ELM, which requires $\mathcal{O}(M^3)$ operations and is typically executed offline, SPLR-ELM performs neuron-wise online updates with $\mathcal{O}(M)$ operations per misclassified sample. This replaces the costly inversion with lightweight local updates, preserving the predictive essence of PLR while eliminating the need for continuous dynamics, eligibility traces, and global modulatory signals. The resulting rule is scalable and hardware-efficient, well-suited for real-time and FPGA-based edge learning.

\section{Accuracy Evaluation}
MNIST~\cite{MINST_DATASET} and Fashion-MNIST~\cite{FASHION_MINST_DATASET} datasets are employed to evaluate the accuracy of SPLR-ELM, enabling fair comparison with existing learning rules on identical benchmarks. A subset of 5,000 training and 1,000 test samples (10\% of the total dataset) is used, yet the model converges stably, demonstrating efficient learning from limited data. Conventional ELM~\cite{ELM} and OS-ELM~\cite{OS_ELM} are included for comparison regarding computational efficiency and accuracy. To ensure fairness, both ELM and OS-ELM employ the \textit{sigmoid} activation and its regularised variant to maintain accuracy and numerical stability. Fig.~\ref{fig:Bar_chart} compares the performance across four representative ELM training paradigms—matrix inversion–based ELM~\cite{ELM}, online sequential ELM~\cite{OS_ELM}, STDP-driven ELM~\cite{STDP_ELM_FPGA}, and the proposed SPLR-based training—evaluated under 32-bit floating-point (FLP32) and 16-bit fixed-point (FXP16, 8 integer + 8 fractional bits) formats. The network architecture is kept consistent, with 784 input nodes, 2048 hidden nodes, and 10 output nodes, and reported results represent average accuracies over training and test datasets. The MNIST, augmented with Gaussian noise to emulate real-world cases, and skewed F-MNIST datasets are introduced to evaluate robustness under non-ideal conditions. The manipulated imbalanced dataset follows a long-tailed class distribution, where the sample count gradually decreases from 400 in the majority class (Class-0) to 200 in the minority class (Class-9), representing a 2× imbalance ratio across classes. An additional experiment using a real-world ten-gesture photonic sensor dataset \cite{zang2024spiking} is included to validate the algorithm’s applicability, featuring 28×28 images with blurrier boundaries than MNIST and F-MNIST. Results on the skewed-F-MNIST datasets present slight accuracy degradation, yet close to an ideal balanced scenario.
\begin{figure}[t]
    \centering
    \includegraphics[width=0.8\linewidth]{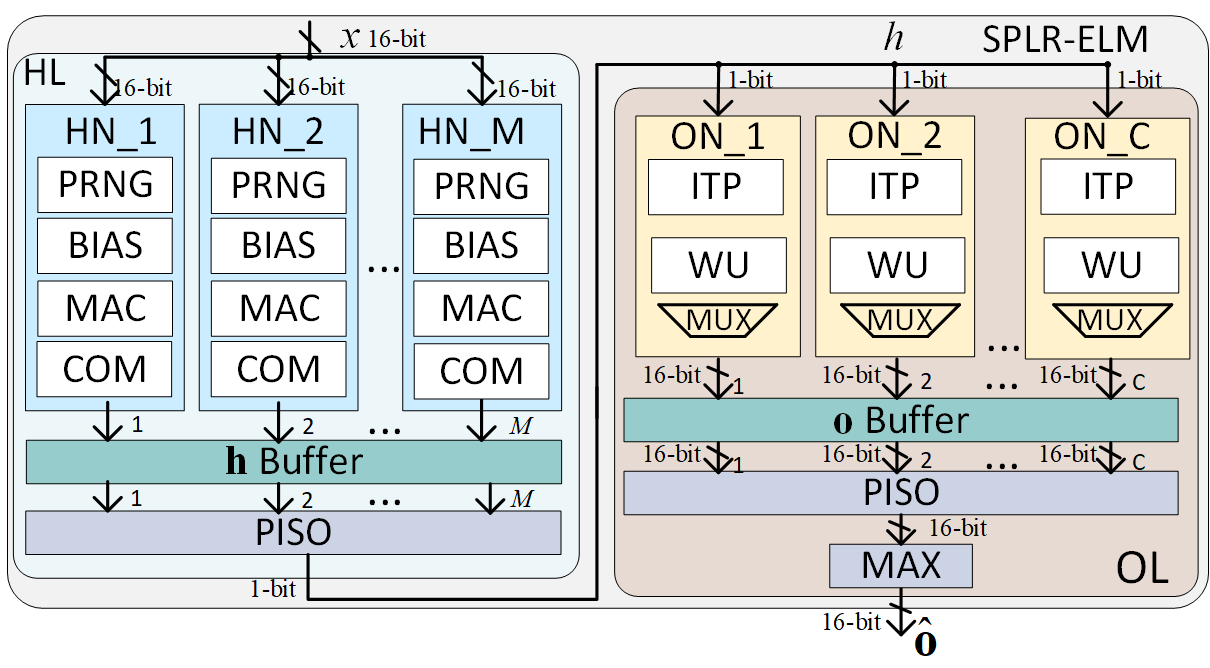}
    \caption{Overview of the proposed SPLR-ELM hardware architecture.}
    \label{fig:top_level}
\end{figure}

Although the latest STDP method (R-STDP)~\cite{STDP_ELM_FPGA}, which shows a similar ELM model architecture but with a different weights updating method, reports higher accuracy in its original paper, its performance degrades in our evaluation due to the use of real-valued pixel inputs, as opposed to the encoded spike trains presented in~\cite{STDP_ELM_FPGA}. SPLR is an error-driven alternative to spike-based R-STDP. While R-STDP adjusts synaptic weights based on spike timing and reward-gated triplet interactions, SPLR relies on the discrepancy between predicted and target outputs, making it computationally and hardware efficient. The SPLR omits spike-timing traces, but preserves the reward-modulated weight adaptation principle. Weight update requires four learning rate hyperparameters to capture short-term and long-term spike-timing effects, making training more stable, Whereas SPLR uses only a learning rate. From a hardware perspective, each hidden node in SPLR only has one comparator and two adders, achieving better scalability than R-STDP, which needs two multipliers. SPLR achieves accuracy comparable to ELM and OS-ELM, while significantly reducing computational complexity. OS-ELM presents slightly lower training accuracy, possibly due to the number of samples in the initial training session being a hyperparameter. The model may not generalise well if the initial batch is small or imbalanced. Although fine-tuning this hyperparameter and diversifying the dataset could improve accuracy, these aspects are beyond the scope of this work. In contrast, SPLR avoids this issue entirely, as its per-sample, prediction error-driven updates require no initial training phase. Even in the FXP16 implementation, the drop in accuracy is slight, further supporting its suitability for hardware-efficient implementations.

\section{Hardware Implementation}
SPLR-ELM is implemented in Verilog using Vivado 2018.3. ZCU104 Ultrascale+ MPSoC, which offers moderate hardware resources, is selected as the target device to demonstrate the feasibility of programmable edge devices. Each HL neuron on the FPGA includes a pseudo-random number generator (PRNG) to generate its input-to-hidden weights at runtime, eliminating the need to preload weights into BRAM. A fixed seed is used during training and inference, and the PNGs are identically reseeded at the arrival of each new sample, ensuring deterministic and consistent weight regeneration across training and inference. This approach significantly reduces BRAM usage by avoiding storage of static weight matrices. Each HL neuron independently generates a unique pseudo-random weight vector of length $D$ (input dimension) for each training or inference sample. 
\begin{figure}[t]
    \centering
    \includegraphics[width=0.75\linewidth]{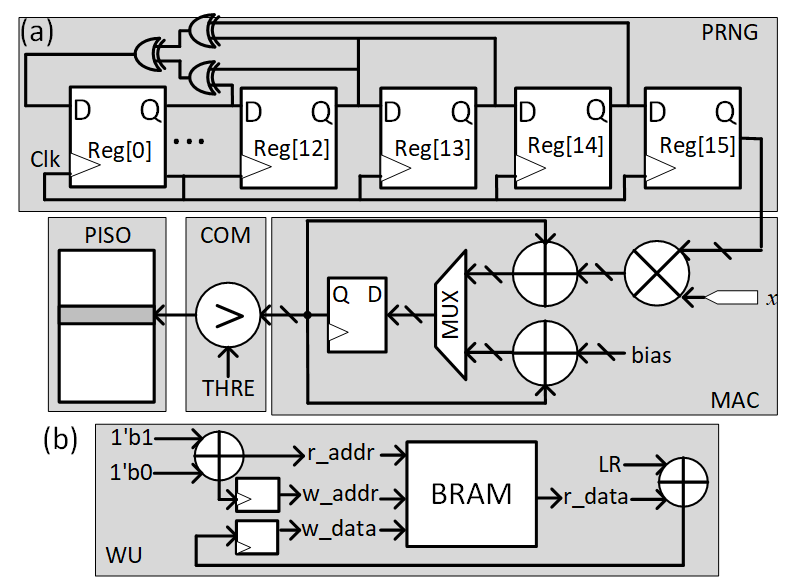}
    \caption{Detailed diagrams of (a) PRNG, MAC, COMP, PISO in a single HN and (B) WU in a single ON. }
    \label{fig:detailed}
\end{figure}

\begin{figure}[t]
    \centering
    \includegraphics[width=0.8\linewidth]{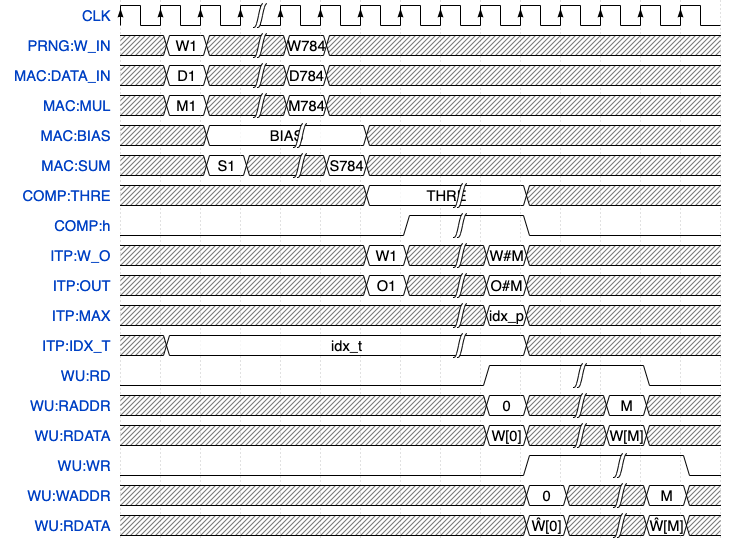}
    \caption{Timing diagram of key modules, including PRNG, MAC, COMP, ITP, and WU.}
    \label{fig:waveform}
\end{figure}

\begin{table}[t]
    \centering
    \begin{threeparttable}
    \caption{Accuracy and FPGA Resource Utilisation for Different Model Sizes}
    \label{tab:fpga_results}
    {\fontsize{6.5}{6.5}\selectfont
    \begin{tabular}{@{}lcccccccc@{}}
    \toprule
    \textbf{\#HN} & 
    \makecell{\textbf{Acc.}\\\textbf{(M)\textsuperscript{a}}} & 
    \makecell{\textbf{Acc.}\\\textbf{(F)\textsuperscript{a}}} & 
    \makecell{\textbf{LUT}\\(230,400)} & 
    \makecell{\textbf{DFF}\\(460,800)} & 
    \makecell{\textbf{DSP}\\(1,728)} & 
    \makecell{\textbf{BRAM}\\(312)} & 
    \makecell{\textbf{Freq.}\\(MHz)} &
    \makecell{\textbf{Power}\\(W)} \\
    \midrule
    512  & $81.3\%$ & $71.2\%$ & 62,864 & 48,543 & 512 & 5 & 230.7 & 1.24 \\
    1024 & $84.1\%$ & $77.2\%$ & 124,690 & 96,162 & 1,024 & 5 & 225.4 & 2.51\\
    1700\textsuperscript{b} & $86.4\%$ & $79.3\%$ & 205,258 & 158,049 & 1,700 & 5 & 224.0 & 3.12 \\
    \bottomrule
    \end{tabular}
    } 
    \begin{tablenotes}
    \item[a] M = MNIST, F = Fashion-MNIST.
    \item[b] Resource-constrained to fit within ZCU104 FPGA fabric.
    \end{tablenotes}
    \end{threeparttable}
\end{table}

\begin{table*}[t]  
\centering
\caption{Performance Comparison of Hardware Implementations Based on Synaptic Plasticity}
\footnotesize
\scriptsize  
\renewcommand{\arraystretch}{0.65}
\label{tab:existing_fpga_compare}
\begin{tabularx}{\textwidth}{
  @{} l l 
  c c c c c
  >{\centering\arraybackslash}X 
  >{\centering\arraybackslash}X 
  >{\centering\arraybackslash}X 
  >{\centering\arraybackslash}X 
  @{}
}
\toprule
\textbf{Work} & \textbf{Platform} & 
\makecell{\textbf{Operating}\\\textbf{freq.}\\\textbf{(MHz)}} & 
\makecell{\textbf{Learn.}\\\textbf{speed}\\\textbf{(fps)}} & 
\makecell{\textbf{Infer.}\\\textbf{speed}\\\textbf{(fps)}} & 
\makecell{\textbf{On-chip}\\\textbf{learning}} & 
\makecell{\textbf{Power}\\\textbf{efficiency}\\\textbf{(fps/W)}}&
\textbf{Learning rule} & 
\textbf{Dataset} & \makecell{\textbf{Network}\\\textbf{architecture}}& 
\textbf{Accuracy (\%)} \\
\midrule
2015 \cite{diehl2015unsupervised} & CPU & N/A & N/A & N/A & N/A & N/A & Pair-based STDP & MNIST & 784-100 & 82.9 \\
\midrule
2021 \cite{wang2021compsnn} & CPU & N/A & N/A & N/A  & N/A& N/A &Tempotron &\makecell{ MNIST, F-MNIST} & 784-100-100 & \makecell{91.22, 77.34} \\
\midrule
2014 \cite{neil2014minitaur} & Spartan-6 FPGA & 75 & N/A & 1.89 & No & 1.26 & Persistent CD & MNIST & 784-500-500-10 & 92.0 \\
\midrule
2017 \cite{wang2017energy} & Virtex-6 FPGA & 120 & 0.06 & 0.12 & Yes  & N/A & Pair-based STDP & MNIST & 784-800 & 89.1 \\
\midrule
2017 \cite{ma2017darwin} & Spartan-6 FPGA & 25 & N/A & 6.25 & No & 52.08 & Persistent CD & MNIST & 784-500-500-10 & 93.8 \\
\midrule
2018 \cite{yousefzadeh2018practical} & Spartan-6 FPGA & 100 & N/A & N/A & Yes  & N/A & Stochastic-STDP & MNIST & 784-6400-10 & 95.7 \\
\midrule
2021 \cite{li2021fast} & Virtex-7 FPGA & 100 & 61 & 317 & Yes  & 196.89 & Pair-based STDP & MNIST & 784-200-100-10 & 92.93 \\
\midrule
2022 \cite{STDP_ELM_FPGA} & ZC706 & 200 & 22.5 & 30 & Yes  & N/A & \makecell{Triplet R-STDP} & \makecell{MNIST, F-MNIST} & 784-2,048-100 & \makecell{93.0, 78.5}\\
\midrule
{2025 This work} & ZCU104 & \textbf{224.0} & \textbf{63,454} & \textbf{122,336} & \textbf{Yes}  & \textbf{39,210.26} & \textbf{SPLR-ELM} & \makecell{MNIST, F-MNIST} & \textbf{784-1,700-10} & \makecell{86.4, \textbf{79.3}} \\

\bottomrule
\end{tabularx}
\end{table*}

The top-level hardware architecture, shown in Fig.~\ref{fig:top_level}, consists of modules for multiplication and accumulation (MAC), binary activation comparison (COMP) as defined in Eq.~\ref{eq:hardset}, a PRNG, in-training prediction (ITP) for computing $\mathbf{o} = \mathbf{h}^\top \mathbf{W}$, and weight update (WU) logic implementing Eq.~\ref{eq:WU}. The hidden (HL) and output (OL) layers instantiate multiple neurons for parallel computation. Once $\mathbf{h}$ and $\mathbf{o}$ are generated, parallel-in-serial-out (PISO) modules aggregate their outputs from the HL and OL, respectively. Multiplexers in the OL toggle between training and inference modes through the ITP module—updating weights during training or outputting $\mathbf{o}$ during inference.Fig.~\ref{fig:detailed}(a) presents the detailed HN structure. The PRNG employs a linear feedback shift register (LFSR) with XOR-based combinational logic to generate pseudo-random weights for the multiplier, incorporating runtime underflow and overflow control. A multiplexer selects between accumulated products or bias summation, followed by threshold comparison to produce binary outputs stored in PISO registers corresponding to neuron indices. The WU module, illustrated in Fig.~\ref{fig:detailed}(b), updates weights stored in BRAM. If the ITP output differs from the ground truth (GT) label, as defined in Eq.~\ref{eq:WU}, the two ONs indexed by $y$ and $\hat{y}$ respectively add and subtract the learning rate (LR) from their stored weights over $M$ clock cycles (CCs).

Fig.~\ref{fig:waveform} presents the timing diagram of a single HN and ON neuron operation, including PRNG and MAC. This neuron is instantiated $M$ times to enable parallel processing of serial input data. After $D$ CCs (784 for the datasets used), $M$ binary activations \textbf{h} are obtained. An additional $M$ CCs are needed to compute the linear output via ITP. If the predicted class $\hat{y}$ differs from the target $y$, the ONs at the indices $\hat{y}$ and $y$ are updated in the $M$ CC according to Eq.~\ref{eq:WU}, using the BRAM access and arithmetic units. Thus, the worst-case total number of CCs required to process one training sample is $T_{\text{cc}_{t}} = D + M + M + P$, and for one inference sample is $T_{\text{cc}_{i}} = D + M + P$, where $P$ is the number of cycles for internal data pipelining (=3 in our case). In the worst case, the second $M$ in $T_{\text{cc}_{t}}$ indicates that all predictions are incorrect. Therefore, the frame-per-second (FPS) rate of the pipelined architecture can be calculated as $\text{FPS} = \frac{f_{\text{max}}}{T_{\text{cc}}}$, where $f_{\text{max}}$ is the maximum clock frequency calculated from timing closure.

The training frame rates (FPS) for the three model sizes in Table~\ref{tab:fpga_results} are 199.7k, 195.1k, and 63.5k, respectively. Compared with prior works summarised in Table~\ref{tab:existing_fpga_compare}, the proposed design achieves substantially higher training and inference throughput, albeit with increased hardware utilisation. The SPLR implementation achieves high computational efficiency through massive neuron-level parallelism and the elimination of multipliers and nonlinear operations, thereby reducing hardware usage and toggle rate and enabling higher throughput per watt. Unlike previous implementations that process encoded spike trains, SPLR-ELM directly handles real-valued inputs, fully leveraging the simplicity of its learning rule to achieve high-speed operation. Table~\ref{tab:fpga_results} also summarises the hardware performance and classification accuracy of SPLR-ELM-FXP16. Power estimates are obtained using the Xilinx Power Estimator, and the maximum operating frequency is extracted from post-implementation timing analysis. Power consumption can be further optimised through time-division multiplexing, lower-bit fixed-point representations for less accuracy-sensitive applications, and clock gating or dynamic adjustment for varying workloads. It is noticeable that SPLR’s accuracy on MNIST is lower than existing works, primarily because we did not adopt time-resolved synaptic state variables during training but used a single-timestamp discretised spike, causing some accuracy loss. Moreover, ELM typically achieves 92--95\% accuracy on MNIST with 2,000--10,000 neurons, while our model reached 86.4\% with 1,700 neurons---only about 5.6\% lower than the 2,000-neuron case, which is acceptable given the trade-off.

\section{Conclusion}
The proposed simplified predictive local rule integrated with ELM provides a compact training strategy that supports on-device learning and inference for image classification. The proposed architecture is implemented on an Xilinx ZCU104 FPGA. Experimental results validate the SPLR-ELM's low latency and competitive accuracy potential on MNIST and Fashion-MNIST datasets. This makes it well-suited for low-complexity classification tasks, particularly on edge devices. Future improvements may focus on extending the algorithm to handle more challenging classification tasks, such as those with low signal-to-noise ratios or high-dimensional data. Additionally, a 2D convolutional SPLR-ELM variant could be explored for enhanced performance in image processing applications.

\bibliographystyle{ieeetr}  
\bibliography{refs}  

@article{PLR_NC_2023,
	title={Sequence anticipation and spike-timing-dependent plasticity emerge from a predictive learning rule},
	author={Saponati, M. and Vinck, M.},
	journal={Nature Communications},
	volume={14},
	number={1},
	pages={4985},
	year={2023},
	publisher={Nature Publishing Group}
}

@article{ELM,
	title={Extreme learning machine: theory and applications},
	author={Huang, G.-B. and Zhu, Q.-Y. and Siew, C.-K.},
	journal={Neurocomputing},
	volume={70},
	number={1--3},
	pages={489--501},
	year={2006},
	publisher={Elsevier}
}

@article{STDP_ELM_FPGA,
	title={A low-cost FPGA implementation of spiking extreme learning machine with on-chip reward-modulated STDP learning},
	author={He, Z. and Shi, C. and Wang, T. and Wang, Y. and Tian, M. and Zhou, X. and Li, P. and Liu, L. and Wu, N. and Luo, G.},
	journal={IEEE Transactions on Circuits and Systems II: Express Briefs},
	volume={69},
	number={3},
	pages={1657--1661},
	year={2021},
	publisher={IEEE}
}

@article{FPGA_PLR_SNN_FPGA,
	title={SC-PLR: An approximate spiking neural network accelerator with on-chip predictive learning rule},
	author={Liu, W. and Xiao, S. and Liu, Y. and Yu, Z.},
	journal={IEEE Transactions on Biomedical Circuits and Systems},
	volume={18},
	number={5},
	pages={1156--1165},
	year={2024},
	publisher={IEEE}
}

@article{MINST_DATASET,
	title={Gradient-based learning applied to document recognition},
	author={LeCun, Y. and Bottou, L. and Bengio, Y. and Haffner, P.},
	journal={Proceedings of the IEEE},
	volume={86},
	number={11},
	pages={2278--2324},
	year={1998},
	publisher={IEEE}
}

@article{FASHION_MINST_DATASET,
	title={Fashion-MNIST: a novel image dataset for benchmarking machine learning algorithms},
	author={Xiao, H. and Rasul, K. and Vollgraf, R.},
	journal={arXiv preprint arXiv:1708.07747},
	year={2017}
}

@article{ELM_ONLINE_HLS,
	title={A neural network-based on-device learning anomaly detector for edge devices},
	author={Tsukada, M. and Kondo, M. and Matsutani, H.},
	journal={IEEE Transactions on Computers},
	volume={69},
	number={7},
	pages={1027--1044},
	year={2020},
	publisher={IEEE}
}

@article{OS_ELM,
	title={Ensemble of online sequential extreme learning machine},
	author={Lan, Y. and Soh, Y. C. and Huang, G.-B.},
	journal={Neurocomputing},
	volume={72},
	number={13--15},
	pages={3391--3395},
	year={2009},
	publisher={Elsevier}
}

@article{OS_ELM_FORGOTTEN,
	title={Online sequential extreme learning machine with forgetting mechanism},
	author={Zhao, J. and Wang, Z. and Park, D. S.},
	journal={Neurocomputing},
	volume={87},
	pages={79--89},
	year={2012},
	publisher={Elsevier}
}

@article{lecun2015deep,
	title={Deep learning},
	author={LeCun, Y. and Bengio, Y. and Hinton, G.},
	journal={Nature},
	volume={521},
	number={7553},
	pages={436--444},
	year={2015},
	publisher={Nature Publishing Group}
}

@article{rumelhart1986learning,
	title={Learning representations by back-propagating errors},
	author={Rumelhart, D. E. and Hinton, G. E. and Williams, R. J.},
	journal={Nature},
	volume={323},
	number={6088},
	pages={533--536},
	year={1986},
	publisher={Nature Publishing Group}
}

@article{RVFL2023,
	title={Random vector functional link network: Recent developments, applications, and future directions},
	author={Malik, Ashwani Kumar and Gao, Ruobin and Ganaie, MA and Tanveer, Muhammad and Suganthan, Ponnuthurai Nagaratnam},
	journal={Applied Soft Computing},
	volume={143},
	pages={110377},
	year={2023},
	publisher={Elsevier}
}

@article{STDP2008,
	title={Spike timing--dependent plasticity: a Hebbian learning rule},
	author={Caporale, Natalia and Dan, Yang},
	journal={Annu. Rev. Neurosci.},
	volume={31},
	number={1},
	pages={25--46},
	year={2008},
	publisher={Annual Reviews}
}

@article{yousefzadeh2018practical,
  title={On practical issues for stochastic STDP hardware with 1-bit synaptic weights},
  author={Yousefzadeh, Amirreza and Stromatias, Evangelos and Soto, Miguel and Serrano-Gotarredona, Teresa and Linares-Barranco, Bernab{\'e}},
  journal={Frontiers in neuroscience},
  volume={12},
  pages={665},
  year={2018},
  publisher={Frontiers Media SA}
}

@article{li2021fast,
  title={A fast and energy-efficient SNN processor with adaptive clock/event-driven computation scheme and online learning},
  author={Li, Sixu and Zhang, Zhaomin and Mao, Ruixin and Xiao, Jianbiao and Chang, Liang and Zhou, Jun},
  journal={IEEE Transactions on Circuits and Systems I: Regular Papers},
  volume={68},
  number={4},
  pages={1543--1552},
  year={2021},
  publisher={IEEE}
}

@article{neil2014minitaur,
  title={Minitaur, an event-driven FPGA-based spiking network accelerator},
  author={Neil, Daniel and Liu, Shih-Chii},
  journal={IEEE transactions on very large scale integration (VLSI) systems},
  volume={22},
  number={12},
  pages={2621--2628},
  year={2014},
  publisher={IEEE}
}

@article{wang2017energy,
  title={Energy efficient parallel neuromorphic architectures with approximate arithmetic on FPGA},
  author={Wang, Qian and Li, Youjie and Shao, Botang and Dey, Siddhartha and Li, Peng},
  journal={Neurocomputing},
  volume={221},
  pages={146--158},
  year={2017},
  publisher={Elsevier}
}

@article{ma2017darwin,
  title={Darwin: A neuromorphic hardware co-processor based on spiking neural networks},
  author={Ma, De and Shen, Juncheng and Gu, Zonghua and Zhang, Ming and Zhu, Xiaolei and Xu, Xiaoqiang and Xu, Qi and Shen, Yangjing and Pan, Gang},
  journal={Journal of systems architecture},
  volume={77},
  pages={43--51},
  year={2017},
  publisher={Elsevier}
}

@article{diehl2015unsupervised,
  title={Unsupervised learning of digit recognition using spike-timing-dependent plasticity},
  author={Diehl, Peter U and Cook, Matthew},
  journal={Frontiers in computational neuroscience},
  volume={9},
  pages={99},
  year={2015},
  publisher={Frontiers Media SA}
}

@article{wang2021compsnn,
  title={CompSNN: A lightweight spiking neural network based on spatiotemporally compressive spike features},
  author={Wang, Tengxiao and Shi, Cong and Zhou, Xichuan and Lin, Yingcheng and He, Junxian and Gan, Ping and Li, Ping and Wang, Ying and Liu, Liyuan and Wu, Nanjian and others},
  journal={Neurocomputing},
  volume={425},
  pages={96--106},
  year={2021},
  publisher={Elsevier}
}

@article{roy2019towards,
  title={Towards spike-based machine intelligence with neuromorphic computing},
  author={Roy, Kaushik and Jaiswal, Akhilesh and Panda, Priyadarshini},
  journal={Nature},
  volume={575},
  number={7784},
  pages={607--617},
  year={2019},
  publisher={Nature Publishing Group UK London}
}

@ARTICLE{CerebelluMorphic2020,
  author={Yang, Shuangming and Wang, Jiang and Zhang, Nan and Deng, Bin and Pang, Yanwei and Azghadi, Mostafa Rahimi},
  journal={IEEE Transactions on Neural Networks and Learning Systems}, 
  title={CerebelluMorphic: Large-Scale Neuromorphic Model and Architecture for Supervised Motor Learning}, 
  year={2022},
  volume={33},
  number={9},
  pages={4398-4412},
  doi={10.1109/TNNLS.2021.3057070}}

@ARTICLE{Fault_Tolerant2022,
  author={Yang, Shuangming and Wang, Jiang and Deng, Bin and Azghadi, Mostafa Rahimi and Linares-Barranco, Bernabe},
  journal={IEEE Transactions on Neural Networks and Learning Systems}, 
  title={Neuromorphic Context-Dependent Learning Framework With Fault-Tolerant Spike Routing}, 
  year={2022},
  volume={33},
  number={12},
  pages={7126-7140},
  doi={10.1109/TNNLS.2021.3084250}}

@ARTICLE{BiCoSS2022,
  author={Yang, Shuangming and Wang, Jiang and Hao, Xinyu and Li, Huiyan and Wei, Xile and Deng, Bin and Loparo, Kenneth A.},
  journal={IEEE Transactions on Neural Networks and Learning Systems}, 
  title={BiCoSS: Toward Large-Scale Cognition Brain With Multigranular Neuromorphic Architecture}, 
  year={2022},
  volume={33},
  number={7},
  pages={2801-2815},
  doi={10.1109/TNNLS.2020.3045492}}

@ARTICLE{Scalable2020,
  author={Yang, Shuangming and Deng, Bin and Wang, Jiang and Li, Huiyan and Lu, Meili and Che, Yanqiu and Wei, Xile and Loparo, Kenneth A.},
  journal={IEEE Transactions on Neural Networks and Learning Systems}, 
  title={Scalable Digital Neuromorphic Architecture for Large-Scale Biophysically Meaningful Neural Network With Multi-Compartment Neurons}, 
  year={2020},
  volume={31},
  number={1},
  pages={148-162},
  doi={10.1109/TNNLS.2019.2899936}}

@article{zang2024spiking,
  title={Spiking Neural Network Enhanced Hand Gesture Recognition Using Low-Cost Single-photon Avalanche Diode Array},
  author={Zang, Zhenya and Li, Xingda and Li, David Day Uei},
  journal={arXiv preprint arXiv:2402.05441},
  year={2024}
}

\end{document}